\documentclass[a4paper,11pt]{article}
\usepackage{latexsym}
\usepackage{amsmath}
\usepackage{amssymb}
\usepackage{bm}
\usepackage[dvipdf,dvips]{graphics}
\usepackage{color}
\title{Mixed Quark-Gluon condensate at finite temperature and density in the global color
symmetry model }
\footnotetext{E-mail:zhangforever@sohu.com }
\author{$\mbox{Zhao Zhang}^{1}$, $\mbox{Wei-qin
Zhao}^{1,2}$\\[5pt] \textit{${}^1$Institute of High Energy
Physics, Academica Sinica,}\\ \textit{Beijing, 100039, P. R.
China}\\
 \textit{${}^2$CCAST(World Laboratory), P.O. Box 8730,}\\
\textit{Beijing 100080, P. R. China}}
\date{}
\begin{document}
\maketitle
\begin{abstract}
The mixed quark-gluon condensate is another chiral order parameter
in QCD, which plays an important role in the application of QCD
sum rules. In this letter, we study the properties of quark-gluon
mixed condensate at finite temperature and quark chemical
potential in the framework of global color symmetry model. Using
an infrared-dominant model gluon two-point function, we find that
the behavior of quark-gluon mixed condensate at finite temperature
and chemical potential is similar to that of the quark condensate,
and both of them give the same information about chiral phase
transition.  We also find that the ratio of these two condensates
is insensitive to the temperature and chemical potential, which
supports the conclusion obtained recently by the authors using
quenched lattice QCD ( They only studied the nature of the mixed
quark-gluon  condensate at finite temperature ).
\\[4pt] {\textit{Keywords}}:\,
Field theory at finite temperature and chemical potential; Vacuum condensate;
 Confinement; Dynamical chiral symmetry breaking; Dyson-Schwinger
equations.\\
PACS number(s): 11.10.Wx; 12.38.Mh; 12.38.Lg; 24.85.+p
\end{abstract}
\newpage
The mixed quark-gluon condensates
$g\langle\overline{q}\sigma^{\mu\nu}G_{\mu\nu}q\rangle$ is one of
the vacuum expectation values which reflect the nonperturbative
structure of QCD vacuum. This condensate plays an important role
for the low energy phenomena of hadrons in the framework of QCD
sum rules\cite{r1,r2}, where the various condensates in the
operator product expansion (OPE) are taken as input parameters
phenomenologically to reproduce various hadronic properties
systematically. Contrary to the well-known condensates of the
lowest dimension, $\langle{G^2}_{\mu\nu}\rangle$ and
$\langle\overline{q}q\rangle$, the mixed condensate represents the
direct correlation between quarks and gluons in QCD vacuum. Due to
the fact that the mixed condensate plays a relevant role in OPE as
a next-to-leading chiral variant operator, the mixed quark-gluon
condensate can be used as another chiral order parameter of the
second lowest dimension for QCD phase transition.
\par
The previous evaluations of the mixed quark-gluon condensate at
zero temperature and chemical potential were performed in the
framework of QCD sum rule method\cite{r2,r3,r4}, quenched lattice
QCD\cite{r5,r6} and QCD effective models\cite{r7,r8},
respectively. The value of this condensate is still not conclusive
from these studies, and the parameter
$m_0^2=g\langle\overline{q}\sigma_{\mu\nu}G^{\mu\nu}q\rangle/\langle\overline{q}q\rangle$
 determined at the renormalization point $u=1\mbox{GeV}^2$ ranges from $0.5\mbox{GeV}^2$ to $3.7\mbox{GeV}^2$. Recently, the behaviour of
the various low-dimensional condensates at finite temperature and
chemical potential is the subject of intensive researches due to
their direct relation to the QCD phase transition. In contrast to
$\langle\overline{q}q\rangle$, the thermal properties of the mixed
quark-gluon condensate of dimension 5 is still least known.
Therefore it is interesting to investigate the thermal nature of
this mixed condensate for it can act as the chiral order parameter
which is independent of the quark condensation
$\langle\overline{q}q\rangle$. Furthermore, the evaluation of the
mixed quark-gloun condensate at finite temperature $T$ and quark
chemical potential $\mu$ has an impact on the hadron phenomenology
through QCD sum rule.
\par
 So far, the investigation of thermal
behavior of
$g\langle\overline{q}\sigma_{\mu\nu}G^{\mu\nu}q\rangle$ has been
performed by the authors\cite{r5} through the quenched lattice QCD
calculation with Kogut-Susskind fermion(KS) field method, where a
critical temperature $T_c\approx{280}$MeV was obtained for both
the chiral symmetry and deconfinement restoration transitions. In
Ref.\cite{r5}, the obtained ratio between the mixed quark-gluon
condensate and the quark condensate is almost independent of the
temperature even in the very vicinity of $T_c$. Since there is
significant overlap between contemporary Dyson-Schwinger-Equations
(DSEs) studies and the numerical simulation of lattice QCD,  the
DSEs could provide an adjunct to lattice QCD. The truncation that
is accurate in the common domain such as zero temperature and
chemical potential and finite temperature can be used to
extrapolate into the finite chemical potential domain, which is
presently inaccessible in lattice simulations. Therefore, it is
necessary to elucidate the $(T,\mu)$ dependence of the mixed
quark-gluon condensate through DSE-models. In this letter, we will
explore the properties of the mixed condensate  at finite
temperature and chemical potential in the frame work of the global
color symmetry model(GCM).
\par
 As a truncated DSE-model, GCM is a quite successful
four-fermion interaction field theory which can be directly
derived through a truncation of QCD\cite{r10}. This truncated DSE
model has been applied extensively at zero temperature and
chemical potential to the phenomenology of QCD\cite{r11,r12},
including the studies of observables from strong interaction to
weak interaction area. Furthermore, the truncated DSE models also
have made important progress in the studies of strong QCD at
finite temperature and chemical potential\cite{r13}.
\par
 In a Euclidean space formulation, with
${\{\gamma_\mu,\gamma_\nu\}}=2\delta_{\mu\nu}$ and
${\gamma^+_\mu=\gamma_\mu}$, the inverse of dressed-quark
propagator at finite temperature and chemical potential in the
framework of GCM takes the form
\begin{eqnarray}
S^{-1}(\tilde{p}_k)&=&i\vec{\gamma}\cdot{\vec{p}}+i\gamma_4(\omega_k+i\mu)+\Sigma(\tilde{p}_k)\\
&=&i\vec{\gamma}\cdot{\vec{p}}A(\tilde{p}_k)+i\gamma_4(\omega_k+i\mu)C(\tilde{p}_k)+B(\tilde{p}_k),
\end{eqnarray}
where $\tilde{p}_k\equiv(\vec{p}, \tilde{\omega}_k)$,
$\tilde{\omega}_k=\omega_k+i\mu$,  $\omega_k=(2k+1)\pi{T}$ is the
Fermi Matasubara frequency, $\mu$ is the chemical potential and
$\Sigma(\tilde{p}_k)$ is the self-energy term of the dressed quark
propagator. The gap equation for the dressed quark in the chiral
limit is determined by the rainbow DSE
\begin{eqnarray}
(A(\tilde{p}_k)-1){\vec{p}}^2=\frac{8}{3}\int_{l,q}g^2D(\widetilde{p}_k-\widetilde{q}_l)
\frac{A(\tilde{q}_l)\vec{q}\cdot\vec{p}}
{A^2(\tilde{q}_l)\vec{q}^2+C^2(\tilde{q}_l)\tilde{\omega}^2_l+B^2(\tilde{q}_l)}\\
(C(\tilde{p}_k)-1){\tilde{\omega}_k}^2=\frac{8}{3}\int_{l,q}g^2D(\tilde{p}_k-\tilde{q}_l)
\frac{C(\tilde{q}_l){\tilde{p}_k}\cdot\tilde{q}_l}
{A^2(\tilde{q}_l)\vec{q}^2+C^2(\tilde{q}_l)\tilde{\omega}^2_l+B^2(\tilde{q}_l)}\\
B(\tilde{p}_k)=\frac{16}{3}\int_{l,q}g^2D(\tilde{p}_k-\tilde{q}_l)
\frac{B(\tilde{q}_l)}
{A^2(\tilde{q}_l)\vec{q}^2+C^2(\tilde{q}_l)\tilde{\omega}^2_l+B^2(\tilde{q}_l)},
\end{eqnarray}
where
$\int_{l,q}=\beta^{-1}\sum_{l=-\infty}^{\infty}\int\frac{d^3q}{(2\pi)^3}$
and $\beta^{-1}=T$. The term $g^2D(\tilde{p}_k)$ is the effective
dressed-gluon propagator at finite temperature and chemical
potential, which is treated as an input parameter in the quark gap
equation. The complex scalar functions $A(\tilde{p}_k)$,
$B(\tilde{p}_k)$  and $C(\tilde{p}_k)$ satisfy:
$F(\tilde{p}_k)^*=F(\tilde{p}_{-(k+1)})$, $F=A, B, C$, although
not explicitly indicated they are functions only of $\vec{q}^2$
and $\tilde{\omega}^2_k$.
\par
We will use the technique proposed in Ref.\cite{r7} to calculate
the mixed quark-gluon condensate at finite temperature and
chemical potential in the framework of global color symmetry
model. In the context of nonzero temperature, the space-time
integration $\int{d^4x}$ is replaced by the form
$\int_0^{\beta}dx_4\int{d^3x}$ and the continuum integration
$\int{d^4p}$ in the momentum space is replaced by Matasubara
frequency summation
$\beta^{-1}\sum_{k=-\infty}^{\infty}\int{d^3p}$. At the mean field
level, the vacuum expectation of any quark operator of the form
\begin{equation}
O_n\equiv(\overline{q}_{j_1}\Lambda_{j_1i_1}^{(1)}q_{i_1})
(\overline{q}_{j_2}\Lambda_{j_2i_2}^{(2)}q_{i_2})
\cdot\cdot\cdot(\overline{q}_{j_n}\Lambda_{j_ni_n}^{(n)}q_{i_n})
\end{equation}
can be calculated straightforward. Here $\Lambda^{(i)}$ is an
operator in Dirac, flavor and color space. Since the generating
functional for the quark field is a typical Gaussian type
integration at the mean field level in the framework of this
four-fermion interaction model, the expression for the vacuum
expectation of quark operator $O_n$ can be easily
derived\cite{r18,r7} and takes the form
\begin{eqnarray}\label{av}
\langle:O_n:\rangle=(-1)^n\sum_p(-)^p[\Lambda^{(1)}_{j_1i_1}\cdots
\Lambda^{(n)}_{j_ni_n}(S)_{i_1j_{p(1)}}\cdots(S)_{i_nj_{p(n)}}],
\end{eqnarray}
where $\Pi$ stands for a permutation of the $n$ indices and $S$ is
the dressed-quark propagator in the mean field level. For example,
we can get the familiar expression for the quark condensate
$\langle\overline{q}q\rangle_T^{\mu}$ at temperature $T$ and
chemical potential $\mu$
\begin{eqnarray}
\label{QQCTU}\langle\overline{q}q\rangle_T^\mu&=&(-)Tr_{DC}{S(x,x)}\nonumber\\
&=&(-)4N_c\int_{l,q}\frac{B(\tilde{q}_l)}{A^2(\tilde{q}_l)\vec{q}^2+C^2(\tilde{q}_l)\tilde{\omega}_l^2+B^2(\tilde{q}_l)}.
\end{eqnarray}
The trace is to be taken in Dirac and color space, whereas the
flavor trace has been separated out. The four quark condensate can
also be obtained easily from  Eq.(\ref{av}) and proved to be
proportional to the square of the quark condensate which is
consistent with the vacuum saturation assumption of Ref.\cite{r7}.
\par
Due to the glounic degree in
$g\langle\overline{q}\sigma_{\mu\nu}G^{\mu\nu}q\rangle$, we can
not get the expression for the mixed quark-gluon condensation
directly from Eq.(\ref{av}). Since the generating functional over
gluon field $A$ in this four-fermion interaction model is
quadratic with a given quark-quark interaction $D$, the
integration over any number of gluon fields can be performed
analytically \cite{r7}. Using the same shorthand notation for the
typical Gaussian integrations as in Ref. \cite{r7}, we have
\begin{eqnarray}
\int\mathcal{D}Ae^{-\frac{1}{2}AD^{-1}A+jA}
&\equiv&{e}^{\frac{1}{2}jDj}\\
\int\mathcal{D}AAe^{-\frac{1}{2}AD^{-1}A+jA}
&\equiv&(jD){e}^{\frac{1}{2}jDj}\\
\int\mathcal{D}AA^2e^{-\frac{1}{2}AD^{-1}A+jA}
&\equiv&[D+{(jD)}^2]{e}^{\frac{1}{2}jDj},
\end{eqnarray}
where $j=\overline{q}\gamma_{\mu}\frac{\lambda^a}{2}q$ is the
quark color current. It should be noted that $D$ stands for the
connected gluon two-point Green function at nonzero temperature
and chemical potential and the integration in the exponent takes
the form $\int_0^{\beta}dx_4\int{d^3x}$ according to the imaginary
time thermal field theory. Due to the transition from the gluon
vacuum average to the quark color current
$\overline{q}\gamma_{\mu}\frac{\lambda^a}{2}q$ together with the
gluon two-point function $D$ which is usually treated as an input
parameter in truncated DSE-model method to get the quark gap
equation, we can use the Eq.(\ref{av}) to evaluate the mixed
quark-gluon condensate at temperature $T$ and chemical potential
$\mu$ in the mean field level . Applying the technique described
above, we obtain
\begin{eqnarray}\label{QGCTUX}
g\langle\overline{q}(x)G_{\mu\nu}(x)\sigma^{\mu\nu}q(x)\rangle&=&
(-2i)N_c\frac{4}{3}\int_z
[\partial_{\mu}^{(x)}g^2D(z-x)]\nonumber\\
&&{\times}tr_D[S(z-x)\sigma_{\mu\nu}S(x-z)\gamma_\nu]\nonumber\\
&&+(4i)N_c\int_{z_1}\int_{z_2} g^2D(z_1-x)g^2D(z_2-x)\nonumber\\
&&{\times}tr_D[S(z_2-x)\sigma_{\mu\nu}S(x-z_1)\gamma_{\mu}S(z_1-z_2)\gamma_\nu],\nonumber\\
\end{eqnarray}
where $x=(x_4,\vec{x})$ and $\int_x$ stands for
$\int_0^\beta{dx_4}\int{d^3x}$.
\par
From the dressed quark gap equation in GCM
\begin{equation}
\Sigma(x-y)=\frac{4}{3}g^2D(x-y)\gamma_{\mu}S(x-y)\gamma_{\mu},
\end{equation}
we can get the form
\begin{equation}
\frac{4}{3}g^2D(x-y)S(x-y)=\frac{1}{4}\frac{1}{4}tr_D[\Sigma(x-y)]-\frac{1}{2}\frac{1}{4}\gamma_{\mu}
tr_D[\Sigma(x-y)\gamma_\mu].
\end{equation}
Using this equation, the integration over the gluon two-point
function $D$ in Eq.(\ref{QGCTUX}) is replaced by the quark
self-energy $\Sigma$. This technique strongly simplifies the
calculation of the mixed condensate. The formula for the
calculation of the mixed quark-gluon condensate is now expressed
only in terms of three scalar functions $A$,$B$ and $C$ which are
solutions of the quark gap equation. After Fourier transformation
and performing the integration over space-time , we get the final
integration expression for the mixed condensate at temperature $T$
and chemical potential $\mu$.
\begin{eqnarray}
\label{QGCTU}g\langle\overline{q}G_{\mu\nu}\sigma^{\mu\nu}q\rangle_T^{\mu}&=&36\int_{k,q}\frac{B[(2-A)\vec{q}^2+(2-C))\tilde{\omega}^2_k]}
{A^2\vec{q}^2+C^2\tilde{\omega}_k^2+B^2}\nonumber\\
&&+\frac{81}{4}\int_{k,q}\frac{B^3+2B[A(A-1)\vec{q}^2+C(C-1)\tilde{\omega}_k^2]}
{A^2\vec{q}^2+C^2\tilde{\omega}_k^2+B^2}.
\end{eqnarray}
Substituting $T=0$ and $\mu=0$ into Eq.(\ref{QGCTU}),  We have
\begin{eqnarray}\label{QGC00}
g\langle\overline{q}G_{\mu\nu}\sigma^{\mu\nu}q\rangle_{T=0}^{\mu=0}&=&(-)
\frac{3}{16\pi^2}\left\{
12\int{ds}s^2B\frac{2-A}{sA^2+B^2}\right.\nonumber\\
&&+\left.\frac{27}{4}\int{ds}sB \frac{[2A(A-1)]s+B^2}
{sA^2+B^2}\right\}.
\end{eqnarray}
This is just the equation obtained in Ref.\cite{r7}. It is clear
according to Eq.(\ref{QGCTU}) that in the Wigner phase
characterized by $B\equiv{0}$, the mixed quark-gluon condensate
takes the value $0$ and the nonzero value of this condensate
signals the nontrivial structure of QCD vacuum. This is the reason
that $g\langle\overline{q}G_{\mu\nu}\sigma^{\mu\nu}q\rangle$ can
also be used as the chiral order parameter as
$\langle\overline{q}q\rangle$ does.
\par
It should be noted that our study ignores effects from hard
gluonic radiative corrections to these condensates due to its
minor importance for our study of nonperturbative properties in
the low and medium energy region. That means the chosen
quark-quark interaction model $g^2D$ has a finite range in
momentum space and the momentum integrals in Eq.(\ref{QGCTU}) are
finite. Because GCM is a no-renormalizable effective field theory,
the scale at which a condensate is defined in our approach is a
typical hadronic scale, which is implicitly determined by the
model gluon propagator $g^2D$ and the solutions of DSE for the
dressed quark gap equations. The similar situation is the
determination of vacuum condensates within the framework of the
instanton liquid model\cite{r21} where the scale is set by the
inverse inatanton size.
\par
To explore the properties of the mixed quark-gluon condensate at
finite temperature and chemical potential, we use the
infrared-dominant model
\begin{equation}\label{GP}
g^2D(\tilde{p}_k)=\frac{3}{16}(2\pi)^3\beta\eta^2\delta_{k0}\delta(\vec{p}),
\end{equation}
which is a generalization to $T\neq{0}$ of the model introduced in
Ref.\cite{r19}, where $\eta\approx{1.06}$GeV is a mass-scale
parameter fixed by fitting $\pi$- and $\rho$-meson masses. As an
infrared enhanced model, it does not represent well the behavior
of $D_{\mu\nu}(\tilde{p}_k)$ away from
$|\vec{p}|^2+\tilde{\omega}_k^2\approx{0}$. Consequently some
model-dependent artifacts arise. However and there is significant
advantage in its simplicity and since the artifacts are easily
identified, the model can exhibits many of the qualitative
features based on the more  sophisticated
A$ns\ddot{a}tze$\cite{r15,r16}. This simple confining  model has
been used successfully to explore the thermal properties of
QCD\cite{r14}.
\par
Substituting Eq.(\ref{GP}) into Eq.(\ref{QGCTU}), we get an
algebraic equation which has two qualitatively distinct solutions.
The Nambu-Goldstone solution, with
\begin{eqnarray}\label{NGA}
A(\tilde{p}_k)=C(\tilde{p}_k)&=&\left\{
       \begin{array}{cc}
             2, &{ \quad \mbox{Re}(\tilde{p}_k^2)<\frac{\eta^2}{4}},\\
              \frac{1}{2}(1+\sqrt{1+\frac{2\eta^2}{\tilde{p}_k^2}}),
               &{\quad \mbox{otherwise}},
       \end{array}\right.\\\label{m32}
\label{NGB}B(\tilde{p}_k)&=&\left\{
       \begin{array}{cc}
             \sqrt{\eta^2-4\tilde{p}_k^2}, & {\quad  \mbox{Re}(\tilde{p}_k^2)<\frac{\eta^2}{4}}, \\
            0, &{\quad
             \mbox{otherwise}},
       \end{array}\right.
\end{eqnarray}
describes a phase of the model in which: $1)$ chiral symmetry is
dynamically broken due to the nonzero quark mass function,
$B(\tilde{p}_k)$, in the chiral limit; and 2) the dressed-quarks
are confined due to the absence of a Lehmann representation of the
dressed quark propagator. The other one is Wigner solution with
for which
\begin{equation}\label{Wigner}
\hat{B}(\tilde{p}_k)\equiv0,\quad
\hat{A}(\tilde{p}_k)=\hat{C}(\tilde{p}_k)=\frac{1}{2}(1+\sqrt{1+\frac{2\eta^2}{\tilde{p}_k^2}}),
\end{equation}
which describes a phase of the model with neither dynamical chiral
symmetry breaking(DCSB) nor confinement.
\par
In Ref.\cite{r14}, this model was used to explore the chiral
symmetry and deconfinement restoration transition between these
two phases characterized by qualitatively different ,
momentum-dependent modification of the quark propagator. The
relative stability of the confined and deconfined phases is
measured by the $(T,\mu)$-dependent vacuum pressure difference or
bag constant\cite{r10}
\begin{equation}
B(T,\mu)\equiv{P[S_{NG}]-P[S_W]},
\end{equation}
where $S_{NG}$ stands for the quark propagator with the
Nambu-Goldstone solution and $S_W$ means the quark propagator with
the Wigner solution. $P[S]$ is the pressure obtained by the
tree-level auxiliary-field effective action in the stationary
phase approximation\cite{r17},  which takes the form
\begin{equation}
P[S]=\frac{T}{V}\{TrLn[\beta{S}^{-1}]-\frac{1}{2}Tr[\Sigma{S}]\}
\end{equation}
 $B(T,\mu)>0$ indicates
the stability of the confined phase and hence the phase boundary
is specified by
\begin{equation}\label{pb}
B(T,\mu)=0.
\end{equation}
In the chiral limit, the deconfinement and chiral symmetry
restoration transition in this model are coincident. Substituting
Eq.(\ref{NGA}-\ref{Wigner}) into Eq.(\ref{pb}), we get the
equation\cite{r14}
\begin{equation}
\frac{2N_cN_f}{\pi^2}\beta^{-1}\sum_{l=0}^{l_{max}}\int_0^{\overline{p}}dpp^2[Re(\frac{2\tilde{p}_k^2}{\eta^2}-{\hat{C}(\tilde{p}_k)}^{-1})-\ln|\frac{\tilde{p}_k^2\hat{C}(\tilde{p}_k)^2}{\eta^2}|]=0
\end{equation}
with $\omega^2_{l_{max}}\leq\frac{\eta^2}{4}+\mu^2$,
$\overline{p}^2=\omega^2_{l_{max}}-\omega^2_l$, which determines
the critical line in the $(T,\mu)$ plane. The critical temperature
for zero chemical potential in this model is $T_c^0=0.170$Gev
which is only $12\%$ larger than the value obtained in
Ref.\cite{r15} using a more sophisticated model and the order of
transition is the same. This illustrates the simple model's
ability to provide a reasonable guide to the thermodynamics
properties of more sophisticated DSE-models. For $T=0$, the
critical chemical potential is $\mu_c^0=0.3$GeV and in case of
nonzero chemical potential the transition is the first
order\cite{r14,r16}.
\par
Substituting Eqs.(\ref{NGA},\ref{NGB}) into Eq.(\ref{QQCTU}) and
Eq.(\ref{QGCTU}), we have the expressions for quark condensate and
mixed quark-gluon condensate  at temperature $T$ and chemical
potential $\mu$ from the infrared-dominant model
\begin{eqnarray}
\label{QQCTU*}-\langle\overline{q}q\rangle=\frac{12}{\pi^2}\frac{T}{\eta^2}\sum_{l=0}^{l_{max}}\int_0^{\tilde{p}}dpp^2Re(\sqrt{\eta^2-4(p^2-\mu^2+{\omega}_l^2+2i\mu\omega_l)}\\
\label{QGCTU*}-g\langle\overline{q}\sigma{G}q\rangle=\frac{81}{4\pi^2}T\sum_{l=0}^{l_{max}}\int_0^{\tilde{p}}dpp^2Re(\sqrt{\eta^2-4(p^2-\mu^2+{\omega}_l^2+2i\mu\omega_l)}).
\end{eqnarray}
Comparing Eq.(\ref{QGCTU*}) with Eq.(\ref{QQCTU*}), we find that
the mixed quark-gluon condensate is proportional to the quark
condensate and the ratio between these two condensates is a
constant independent of temperature $T$ and chemical potential
$\mu$. It is a nontrivial result that the behavior of the mixed
guark-gluon condensate at finite temperature and chemical
potential is similar to the quark condesate, for
$g\langle\overline{q}\sigma{G}q\rangle$ reflects the color-octet
component of quark-antiquark pairs in the QCD vacuum ,while
$\langle\overline{q}q\rangle$ reflects only the color-singlet
quark-antiquark components. That means the mixed condensate
$g\langle\overline{q}\sigma{G}q\rangle$ represents the direct
correlation between color-octet $q-\overline{q}$ pairs and the
gluon field strength  $G_{\mu\nu}$, while
$\langle\overline{q}q\rangle$ represents the direct correlation
between quarks and antiquarks. This nontrivial result was also
obtained in Ref.\cite{r9} by quenched lattice QCD simulation with
the KS-fermion method(note that in Ref.\cite{r9}, the authors only
explored the temperature dependence of the mixed condensate and
the critical temperature for chiral symmetry restoration is 280MeV
 ). The ratio
$m_0^2=g\langle\overline{q}\sigma{G}q\rangle/\langle\overline{q}q\rangle$
is $1.90\mbox{GeV}^2$ in this infrared-dominant model, which
suggests the significance of the mixed condensate in OPE.
\par
In Ref.\cite{r20}, these two condensates take the form
\begin{eqnarray}
\langle\overline{q}q\rangle=\frac{1}{V}\int{d}\lambda\frac{m\rho(\lambda)}{\lambda^2+m^2},\\
g\langle\overline{q}\sigma{G}q\rangle=\frac{1}{V}\int{d}\lambda\frac{m\rho(\lambda)}{\lambda^2+m^2}\langle\lambda|\sigma{G}|\lambda\rangle,
\end{eqnarray}
where $|\lambda\rangle$ stands for the eigenvector of the Dirac
operator as
$i\hat{D}|\lambda\rangle=\lambda|\lambda\rangle$, and
$\rho(\lambda)$ the spectral density on $\lambda$. So the ratio
$m_0^2$  can be expressed as
$\langle\lambda|\sigma{G}|\lambda\rangle_{\lambda=0}$ in the
chiral limit. The insensitive $(T,\mu)$ dependence of $m_0^2$
suggests that the value of
$\langle\lambda|\sigma{G}|\lambda\rangle_{\lambda=0}$ has a weak
dependence on both the temperature and chemical potential in the
Nambu-Goldstone phase, even in the very vicinity of critical
point.
\par

 In Fig.1, the thermal effect on
$-g\langle\overline{q}\sigma{G}q\rangle$ is plotted against
temperature $T$ at several choosen chemical potential points. It
is clearly indicated from Fig.1 that
$-g\langle\overline{q}\sigma{G}q\rangle$ decreases continuously to
zero with increasing $T$. The chemical potential dependence of
$-g\langle\overline{q}\sigma{G}q\rangle$ is shown in Fig.2 and
obviously the mixed condensate increases with $\mu$, up to a
critical value $\mu_c(T)$ when it drops discontinuously to zero.
This behavior is the same as $-\langle\overline{q}q\rangle$, which
can be attributed to the combination $\mu^2-\omega_l^2$ appearing
in Eq.(\ref{QQCTU*}) and Eq.(\ref{QGCTU*}).
\par
We can see from above analysis that though
$g\langle\overline{q}\sigma{G}q\rangle$ and
$\langle\overline{q}q\rangle$ characterize different aspects of
QCD vacuum, both of them not only give the same critical point for
the chiral symmetry restoration,  but also have the same critical
behavior for the QCD phase transition. This conclusion confirms
the result in Ref.\cite{r9} and seems to indicate that all order
parameters for a phase transition have the universal critical
behavior near the critical point.
\par
In summary, we have estimated the $(T,\mu)$ dependence of
$g\langle\overline{q}\sigma{G}q\rangle$ in the framework of GCM.
Using a simple,confined DSE-model of QCD, we get that the value of
the mixed quark-gluon condensate is proportional to the quark
condensate and the ratio between these two condensates is
independent of $T$ and $\mu$. The large ratio
$m_o^2=1.90\mbox{GeV}^2$ suggests that the mixed quark-gluon
condensate plays an important role in QCD sum rules. As two
low-dimensional chiral order parameters, the mixed quark-gluon
condensate and quark condensate give the same critical behavior
for chiral symmetry restoration. As mentioned above, this
infrared-dominant model is poor to represent the ultraviolet
behavior of the gluon propagator, so there may  exist some
model-dependent artifacts in our study. Because this model can
provide a reasonable guide to the thermodynamics properties based
on more sophisticated DSE-models of QCD, we believe that these
results are qualitatively coincident with the more complex
A$ns\ddot{a}tze$ . The more sophisticated investigation is under
progress.

\newpage
\begin{center}{Figures}\end{center}
\vspace{2cm}
\begin{figure}[h]
\begin{center}
\begin{picture}(250,250)
\includegraphics{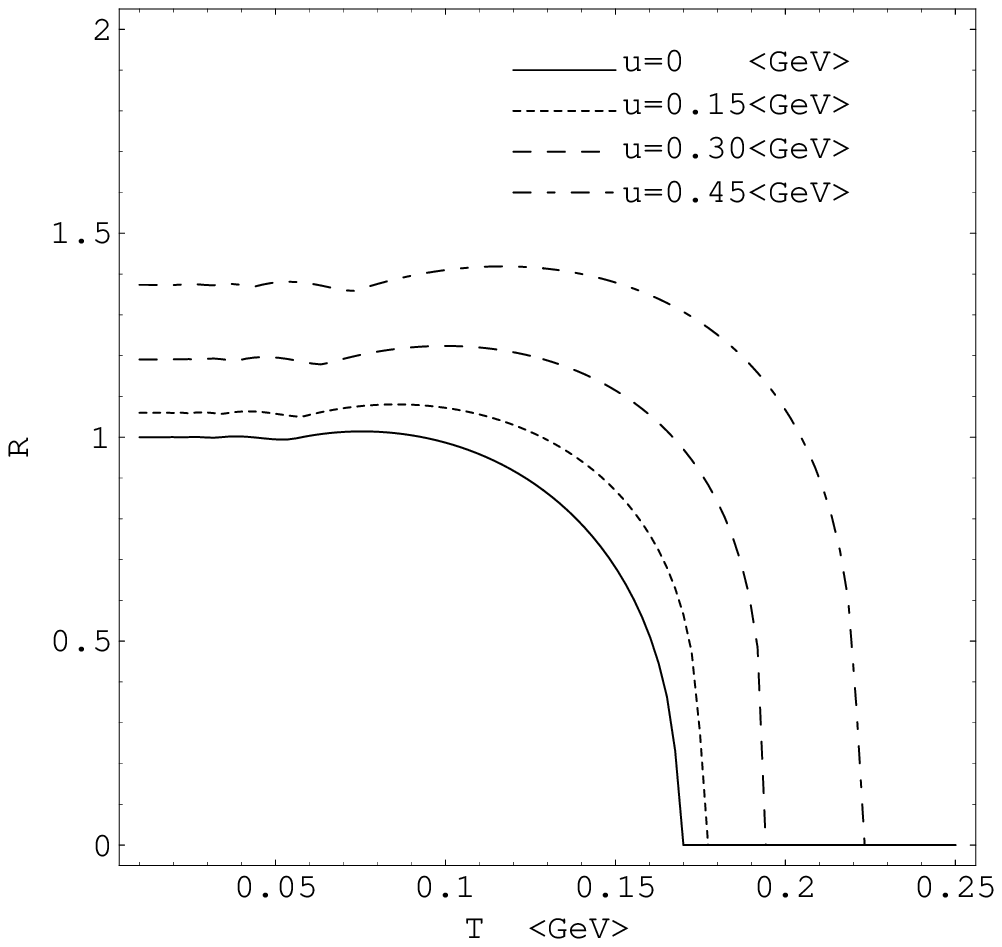}
\end{picture}
\end{center}
\end{figure}
Fig.1. The ratio
$R=(g\langle\overline{q}\sigma{G}q\rangle^\mu_T)^{(\frac{1}{5})}
/(g\langle\overline{q}\sigma{G}q\rangle^{\mu=0}_{T=0})^{(\frac{1}{5})}$,
as a function of $T$ for a range values of $\mu$. The ``structure"
in those curves, apparent for small $T$, is an artifact of the
fact that the infrared-dominant model does not represent well the
the quark-quark interaction in the ultraviolet domain\cite{r14}.
\newpage
\begin{figure}[h]
\begin{center}
\begin{picture}(250,250)
\includegraphics{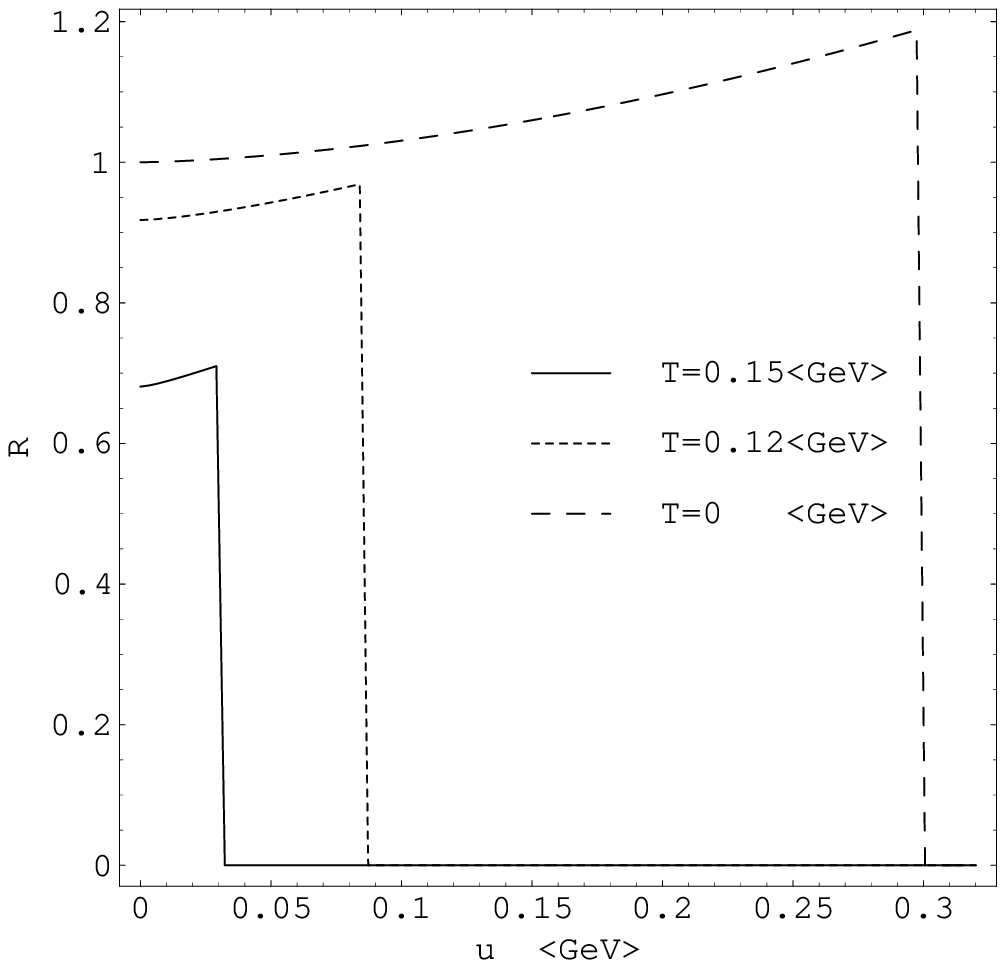}
\end{picture}
\end{center}
\end{figure}
Fig.2.The ratio
$R=(g\langle\overline{q}\sigma{G}q\rangle^\mu_T)^{(\frac{1}{5})}
/(g\langle\overline{q}\sigma{G}q\rangle^{\mu=0}_{T=0})^{(\frac{1}{5})}$,,
as a function of $\mu$ for a range values of $T$.
\end{document}